\documentstyle[12pt]{article}
\begin{document}
\newcommand{\beq}{\begin{equation}}
\newcommand{\eeq}{\end{equation}}
\newcommand{\bea}{\begin{eqnarray}}
\newcommand{\eea}{\end{eqnarray}}
\newcommand{\eps}{\varepsilon}
\newcommand{\Veff}%
{{\cal V}^{\mbox{\scriptsize p}}_{\mbox{\scriptsize eff}}}
\newcommand{\Afr}{A^{\mbox{\scriptsize fr}}}
\newcommand{\Bfr}{B^{\mbox{\scriptsize fr}}}
\newcommand{\Binf}{B^{\mbox{\scriptsize inf}}}
\newcommand{\Blp}{B^{\mbox{\scriptsize LPA}}}
\newcommand{\Fs}{\mbox{\scriptsize F}}
\vskip 5 true cm

\centerline{\bf The local potential approximation for the Brueckner
G-matrix}

\vskip 0.5 cm

\centerline{M.~Baldo$^{a}$, U.~Lombardo$^{b,c}$,
E.E.~Saperstein$^{d}$ and M.V.~Zverev$^{d}$}
\vskip 0.5 cm

\centerline{$^a$INFN, Sezione di Catania, 57 Corso Italia,
I-95129 Catania, Italy}
\centerline{$^b$INFN-LNS, 44 Via S.-Sofia, I-95123 Catania, Italy}
\centerline{$^c$Dipartimento di Fisica, 57 Corso Italia,
I-95129 Catania, Italy}
\centerline {$^{d}$ Kurchatov Institute, 123182, Moscow, Russia }
\vskip 0.5 cm

\centerline{Abstract}
\vskip 0.5 cm

The Brueckner $G$-matrix for a slab of nuclear matter
is analyzed in the singlet $^1S$ and triplet $^3S+{^3D}$ channels.
The complete Hilbert space is split into two domains, the model
subspace $S_0$, in which the two-particle propagator is calculated
explicitly, and the complementary one, $S'$, in which
the local potential approximation is used. This kind of
local approximation was previously found to be quite
accurate for the $^1S$ pairing problem.
A set of
model spaces $S_0(E_0)$ with different values of
the cut-off energy $E_0$ is considered, $E_0$ being
the upper limit for the
single-particle energies of the states belonging to $S_0$.
The independence of the $G$-matrix of $E_0$ is assumed as a criterion
of validity of the local potential approximation. Such independence
is obtained
within few percent
for $E_0{=}10 \div 20\;$MeV for both the
channels under consideration.

\newpage

              {\bf 1. Introduction}.
\vskip 0.2 cm

Most reliable predictions of nuclear properties
come from phenomenological approaches including the
macroscopic-microscopic method \cite{Str,M-S,Tond}, and, at a
more fundamental level,
the Finite Fermi Systems (FFS) theory \cite{AB1,AB2,KhS}, the HF
method with effective forces \cite{VB,HF} and the new versions of the
energy functional method \cite{Fay}. These methods, being comparatively
simple, permit to carry out systematic calculations for a lot of atomic
nuclei.
On the other hand, being phenomenological, these approaches have the
necessity of introducing a set of
adjustable parameters. This point appears especially
delicate when one deals with new types of
nuclear systems, e.g. nuclei in the drip-line vicinity. Indeed,
any extrapolation of phenomenological parameters found for
stable nuclei is hardly reliable in this case.
Therefore the old problem of the {\it ab initio} calculation
of these parameters is of importance not only from an heuristic point
of view but also from the practical one.

Application of methods of microscopic theory of nuclear matter to
finite nuclei (see, e.g., the monographs \cite{Br} and \cite{Schuck})
are mainly limited by use of so-called local density
approximation (LDA). The LDA works reasonably well inside a nucleus
but fails at the surface where there is a domain of density values
for which nuclear matter is unstable. However, it is impossible to
develop a self-consistent nuclear theory without consistent description
of the nuclear surface. Indeed, just in the surface region the
nuclear mean field  sharply changes from zero outside the
nucleus to a constant value inside it.
In terms of the FFS theory, it is associated with a sharp variation of the
scalar-isoscalar Landau-Migdal amplitude $f$ from a strong attraction
outside a nucleus to an almost zero inner value. A
consistent description of this variation is of primary importance for
the nuclear theory. We intend to develop an approach to this
problem based on the Brueckner theory for nonuniform systems
beyond the LDA.

Recently \cite{BLSZ1}, dealing with $^1S$-pairing problem for
semi-infinite nuclear matter, we developed a method of solving
the Bethe-Goldstone (BG) type equation for the separable form of
the Paris $NN$-potential without use of any form of local approximation.
The same method was applied to the case of the slab geometry in Refs.
\cite{BLSZ2,BLSZ3}. Though the separable form of the $NN$-interaction
simplifies calculations significantly, they remain to be rather
cumbersome and much CPU time consuming.
To circumvent this difficulty, we devised
a new version of the local approximation
resulting from the investigation of the effective pairing interaction
$\Veff$ \cite{BLSZ1}. To distinguish
it from the standard LDA, we named it as the local potential approximation
(LPA). The effective interaction is associated with
splitting of the complete Hilbert space  $S$
into two domains. The first one is the model subspace
$S_0$, in which the gap equation is written down in terms of the
effective interaction $\Veff$. The second one is the
complementary subspace $S^{\prime}$, in which the equation
for $\Veff$ is obtained in terms of the free $NN$-interaction
$\cal V$. In this space, the pairing effects are not significant,
therefore the equation for $\Veff$ has the form of
the BG equation.
Dealing with the pairing problem, the model space
was taken in a form convenient for nuclear application, which includes
all the single-particle states with negative energies $\eps_{\lambda}$.
The LPA is related to calculation of the two-particle
propagator in the complementary space which enters the
equation for $\Veff$. Roughly speaking, the LPA procedure
consists in replacing the exact BG propagator by a
suitable form taken from
infinite nuclear matter. For a fixed value of the average
centre-of-mass (CM) $x$-coordinate
$X=(X_{12}+X_{34})$ of incoming and outcoming
nucleons (the $x$-axis is perpendicular to the layer),
the propagator is supposed to be equal to the one
of nuclear matter in the potential well $V(X)$.
Such an approximation turned out to be accurate at a level of a few
percent even at the surface \cite{BLSZ1}. It was shown by a comparison
of the direct solution for $\Veff$ with the LPA one.

From the computational point of view, the problem of finding the
Landau-Migdal interaction amplitude in terms of the Brueckner $G$-matrix
is much more complicated than the pairing problem.
First, the additional triplet $^3S$-channel
(coupled with the $^3D$ one) should be considered for which the
calculations turned out to be
more complicated than those for the singlet $^1S$-channel.
Second, instead of fixing the value of the total perpendicular
momentum $P_{\perp}=0$ as in the pairing problem, the integral over
$P_{\perp}$ appears in this case.
Therefore, though the direct numerical solution of the problem,
in principle, is possible \cite{BLSZ4}, it looks very difficult.
Therefore it is natural to attempt using the LPA for solving the BG
equation for the $G$-matrix that simplifies calculations
significantly. In the case of the singlet $^1S$-channel, the
accuracy of the LPA for the BG equation is
just the same as in the pairing problem because the corresponding
two-particle propagators in the complementary space coincide.
As to the triplet channel, applicability of the LPA is not obvious at all.
The main goal of this paper is to clarify the latter point.

In the case of the BG equation, there is no evident gain from
the introduction of
the effective interaction and it seems more reasonable,
after splitting the Hilbert space as $S = S_0 + S'$,
to formulate the LPA procedure in a direct way.
According to that splitting, the two-particle propagator
$A$ in the BG equation can be written  as a sum of $A=A_0+A'$.
The model space term, $A_0$, should be calculated exactly,
whereas the second one, $A'$, within the LPA.
It is clear that the applicability of the LPA depends on the
choice of the model space $S_0$. Indeed, all the quantum
and finite range effects originate mainly from the states
nearby the Fermi surface whose contribution to the BG equation
is strengthened by the small values of the energy denominator
in the two-particle propagator. These contributions produce the
long-range components of $A$ and should be taken into account exactly.
At the same time,
the individual contribution of a far-lying state
is negligible  and only a sum of a big number of such states
is important. They produce the short-range term of $A$ and can be
considered within the local approximation.
Hence, the accuracy of the LPA should depend on the choice of
the model space: wider $S_0$ is  higher it is.
We use this simple physical idea to impose a criterion of the
applicability of the LPA. We define the model space
$S_0(E_0)$ including all the single particle states with the energies
$\eps_{\lambda} < E_0$ which is more general than that for the pairing
problem \footnote{The latter corresponds to $E_0{=}0$}.
It is obvious that
a larger $E_0$ corresponds to a higher accuracy of the LPA.
We consider the LPA to be
valid at some value of $E_0$ if the $G$-matrix does not practically
change with additional increase of $E_0$.

The paper is organized as follows. Sect.~2 contains the BG
equation for the slab of nuclear matter with separable $NN$-forces.
In Sect.~3  the splitting of the Hilbert space on the
model subspace defined by the cut-off energy $E_0$ and its
complementary subspace
is discussed. In addition, the LPA for the BG equation is introduced.
In Sec.~4 the validity of the LPA for determining the $G$-matrix in the
singlet $^1S$-channel is analyzed for different values of $E_0$.
In Sec.~5 the analysis is extended to the triplet $^3S+{^3D}$-channel.
Sec.~6 contains a summary of the results.

\newpage
\vskip 0.3 cm
              {\bf 2. Bethe-Goldstone equation for the slab system}.
\vskip 0.2 cm

Let us consider the BG equation for the $G$-matrix of two nucleons
at the Fermi surface, i.e. with the single-particle
energies $\eps_{\lambda}=\mu$, where $\mu$ is the chemical potential
of the system under consideration.
In a short notation it reads:
%1
\beq
G(E) = {\cal V} + {\cal V} A(E) G(E),
\label{BG}
\eeq
where  ${\cal V}$ is the free $NN$-potential, $E=2\mu$, and $A$ is the
two-particle propagator which is determined by the integration over the
relative energy of the product $({\cal G}^p {\cal G}^p)$ of two {\it particle}
single-particle Green's functions. The contribution $({\cal G}^h {\cal G}^h)$
of two {\it holes} is neglected.

To speed up the convergence it is convenient to renormalize eq. (\ref{BG})
in terms of the free off-shell $T$-matrix taken at negative
energy $E=2\mu$. The latter obeys the Lippman-Schwinger equation:
%2
\begin{equation}
T(E)={\cal V} + {\cal V} \Afr (E)T(E),
\label{TM}
\end{equation}
where $\Afr (E)$  is the propagator of two free nucleons
with the total energy $E$.

The renormalized BG equation has the form:
%3
\beq
G = T+ T (A - \Afr ) G.
\label{BGT}
\eeq

We consider a nuclear-matter slab of thickness $2L$ placed into
the one-dimensional Saxon-Woods potential $V(x)$ symmetrical with
respect to the point $x=0$
with depth of $V_0=50$ MeV, the diffuseness parameter
$d=0.65$ fm, and $L=8$ fm.

We use the separable version \cite{Par1,Par2}  of the Paris
$NN$-potential \cite{Paris}  which for the
$^1S_0$-channel has the ($3 \times 3$) form:
%4
\begin{equation}
{\cal V}({\bf k},{\bf k}^{\prime}) =
\sum_{ij} \lambda_{ij} g_i(k^2)g_j(k^{\prime 2}).
\label{Par1}
\end{equation}

For the triplet $^3S_1+{^3D_1}$-channel, a similar
$(4 \times 4)$ expansion (\ref{Par1})
is valid with a formal substitution
$g_i(k^2) \to \hat{g}_i(k^2)$ where the column
$\hat{g}_i$ contains two components:
%5
\beq
\hat{g}_i(k^2) =
{ g_i^{L=0}(k^2)  \choose  g_i^{L=2}(k^2)},
\label{Par2}
\eeq
$L$ being the relative orbital moment in the CM system.

The scheme of solving the BG equation for a slab of nuclear system
in the mixed coordinate-momentum representation, which has been
devised in Ref.\cite{BLSZ4}, is adopted also here. Therefore we write down
in the explicit form only
those equations which are necessary for understanding
the procedure and refer
to \cite{BLSZ4} for details. We consider first the singlet channel
$S=0$. As it was mentioned above, all relations remain valid for the
triplet channel as well after replacing
$g_i(k^2) \to \hat{g}_i(k^2)$.

The separable form (\ref{Par1}) of the $NN$-potential in eqs.
(\ref{BG}) and (\ref{TM}) induces similar expansions for the
$G$-matrix
%6
\beq
{G}(k^2_{\perp},k^{\prime 2}_{\perp}, {\bf P_{\perp}};x_1,x_2,x_3,x_4;E)
 =\sum\limits_{ij} G_{ij}(X,X';E, {\bf P_{\perp}})
g_i(k^2_{\perp},x)g_j(k^{\prime 2}_{\perp},x'),
\label{GMij}
\eeq
and $T$-matrix
%7
\beq
{T}(k^2_{\perp},k^{\prime 2}_{\perp}, {\bf P_{\perp}};x_1,x_2,x_3,x_4;E)
 =\sum\limits_{ij} T_{ij}(X{-}X';E, {\bf P_{\perp}})
g_i(k^2_{\perp},x)g_j(k^{\prime 2}_{\perp},x').
\label{TMij}
\eeq

Here the form factors $g_i(k^2_{\perp},x)$ in the mixed
representation are determined by the inverse Fourier transformation
of $g_i(k^2_{\perp}{+}k_x^2)$  with respect to variable $k_x$.
Their analytical form can be found in \cite{BLSZ1} for the singlet
channel and in \cite{BLSZ4}, for the triplet one.
The obvious notation for the CM and relative
coordinates in the $x$-direction are used in (\ref{GMij})
and (\ref{TMij}). Of course, the $T_{ij}$ coefficients depend only
on the difference $t=X{-}X'$ of the CM coordinates.
In the perpendicular direction, the total momentum ${\bf P_{\perp}}$ and
the relative momentum ${\bf k_{\perp}}$ are introduced.

Substitution of (\ref{GMij}) and (\ref{TMij}) into eq.
 (\ref{BGT}) results in a set of one-dimensional integral equations:
%8
$$
G_{ij}(X,X';E,{\bf P}_{\perp}) =T_{ij} (X{-}X';E, {\bf P_{\perp}})+
$$
\beq
 + \sum_{lm} \int dX_1\,dX_2\,T_{il}(X-X_1;E, {\bf P_{\perp}}) \,
\delta B_{lm}(X_1,X_2;E,{\bf P}_{\perp}) \,
G_{mj}(X_2,X';E,{\bf P}_{\perp}),
\label{GMeq}
\eeq
where
%9
\beq
\delta B_{lm} = B_{lm} - \Bfr_{lm}
\label{dB}
\eeq
is the difference between the convolution $B_{lm}$ of the two-particle
propagator $A$ with two form factors $g_l,g_m$ and the
analogous convolution $\Bfr_{lm}$ for the free propagator
$\Afr$. The explicit form of $B_{lm}$ is as follows:
%10
\bea
B_{lm}(X,X';E,{\bf P}_{\perp}) =  \sum_{nn'}
           \int{ d {\bf k}_{\perp} \over {(2\pi)^2} }
{ (1 - n_{\lambda})\,(1 - n_{\lambda'})
\over {E -P_{\perp}^2/4m -\eps_n-\eps_{n'}-k^2_{\perp}/m } }
\times
\nonumber \\
\times
{g_{nn^{\prime}}^l(k^2_{\perp},X)\,
g_{n^{\prime}n}^m (k^2_{\perp},X')}.
\label{Blm}
\eea
 Here we have used a short notation of
$ \lambda = (n, {\bf p}_{\perp})$,
 $\lambda'=(n', {\bf p'}_{\perp})$,
$ {\bf p}_{\perp}= {\bf P}_{\perp}/2 + {\bf k}_{\perp}$,
$ {\bf p'}_{\perp}= {\bf P}_{\perp}/2 - {\bf k}_{\perp}$,
and $n_{\lambda}{=}\theta(\mu {-} \eps_{\lambda})$,
where $\eps_{\lambda}{=}\eps_n{+}p_{\perp}^2/2m$,
and $\eps_n$ are the eigenenergies of the one-dimensional Schr\"odinger
equation with the Saxon-Woods potential. The corresponding
eigenfunctions $y_n(x)$ (they are chosen to be real) enter
the matrix elements of of the form factors:
%11
\begin{equation}
   g_{n,n^{\prime}}^l(k^2_{\perp},X) = \int dx
   \,g_l(k^2_{\perp},x)\,
 y_n(X{+}x/2)  y_{n^{\prime}}(X{-}x/2)
\label{Mel}
\end{equation}
It should be noted that the symbolic sum over
$nn'$ in (\ref{Blm}) actually includes the summation over discrete
states and the integration over the continuum spectrum with the
standard substitution $\sum_n \to \int dp/2\pi $.

In the singlet channel, the BG equation for the G-matrix is very
similar to the one for the effective pairing interaction \cite{BLSZ1}.
Just as in the latter case, it is convenient to extract the
singular $\delta$-form Born term from the complete $G$-matrix:
%12
\beq
G =  {\cal V}  + \delta G.
\label{dG}
\eeq

The equation for the correlation component $\delta G$ of the $G$-matrix
can be readily found from (\ref{BG}):
%13
\beq
 \delta G ={\cal V} A {\cal V}  + {\cal V}  A \delta G.
\label{eqdG}
\eeq

An analogous extraction of the Born term should be made also for the
$T$-matrix:
%14
\beq
T =  {\cal V}  + \delta T.
\label{dT}
\eeq

As a result, the renormalized eq. (\ref{BGT}) yields the following
form:
%15
\beq
 \delta G = F + T (A - \Afr )  \delta G,
\label{eqdG1}
\eeq
where the nonhomogeneous term is
%16
\beq
 F = \delta T + T (A - \Afr ) {\cal V}.
\label{eqdG2}
\eeq

The explicit transformation of eqs.~(\ref{eqdG1}), (\ref{eqdG2}) to
the form similar to eq. (\ref{GMeq}) is quite obvious.
A simplification of the numerical procedure for solving eq.(\ref{eqdG1})
in the slab system under consideration can be made using the
parity conservation which follows from the symmetry of the
Hamiltonian under the inversion transformation $x \to -x$.
As a result, the eigenfunctions $y_n$ can be separated on
the even functions, $y_n^+$, and the odd ones, $y_n^-$. Then
the two-particle propagator in the above equations is the sum
%17
\beq
  A = A^+ + A^-
\label{Apm}
\eeq
of the even and odd components. The first one, $A^+$, originates
from the terms of the sum (\ref{Blm}) containing states ($\lambda,\lambda'$)
with the same parity, and the second one, $A^-$, from those with opposite
parity. So long as  the $NN$-potential ${\cal V}$ does conserve
the parity, the propagators $A^+$ and $A^-$ are not mixed in the BG
equation. Therefore the correlation part of the $G$-matrix is also a sum
of the even and odd components,
%18
\beq
\delta  G =\delta G^+ + \delta G^-,
\label{dGpm}
\eeq
which obey the separated equations:
%19
\beq
\delta  G^{\pi} ={\cal V} A^{\pi} {\cal V} + {\cal V} A^{\pi} \delta G^{\pi},
\label{eqdGpm}
\eeq
$\pi= +, -$.

It is obvious that the integral equation
(\ref{eqdGpm}) can be reduced to the form containing positive $x$ values only
which simplifies the calculation procedure. This equation should be solved
for both values of $\pi$ separately, then the complete $G$-matrix
is found from eqs. (\ref{dG}),(\ref{dGpm}).

All the above general equations remain valid for the triplet
channel $S=1$. The main change occurs in the definition of the
convolution integral (\ref{Blm}).
For the triplet channel it has the form:
%20
\bea
B^{S=1}_{lm}(X,X';E,{\bf P}_{\perp}) =  \sum_{nn'}
           \int{ d {\bf k}_{\perp} \over {(2\pi)^2} }
{ (1 - n_{\lambda})\,(1 - n_{\lambda'})
\over {E -P_{\perp}^2/4m -\eps_n-\eps_{n'}-k^2_{\perp}/m } }
\times
\nonumber \\
\times
{\left(
g^{(0)l}_{nn'}(k^2_{\perp},X)\,
g^{(0)m}_{n'n} (k^2_{\perp},X') +
g^{(2)l}_{nn'} (k^2_{\perp},X)\,
g^{(2)m}_{n'n} (k^2_{\perp},X')
\right)}.
\label{B1lm}
\eea

It should be noted also that all multipole expansions
of eqs. (\ref{GMij}) and (\ref{TMij}) type become the
$2\times 2$ matrices. For example, let us write down
the components of the $G$-matrix in the explicit form:
%21
\beq
%\lefteqn{
G^{LL'}(k^2_{\perp},k^{\prime 2}_{\perp}, {\bf P_{\perp}};x_1,x_2,x_3,x_4;E)=
%}
%\nonumber \\
%& &{}=
\sum\limits_{ij} G_{ij}(X,X';E,{\bf P_{\perp}})\,
g^{(L)}_i(k^2_{\perp},x)g^{(L')}_j(k^{\prime 2}_{\perp},x'),
\label{GLL1}
\eeq
where $L,L'$ equal to 0 or 2.

\vskip 0.3 cm
              {\bf 3. Choice of the model space and the local potential
approximation}.
\vskip 0.2 cm

The main computational problems of solving the BG equation for the
slab system are connected with the calculation of the propagators
(\ref{Blm}) and (\ref{B1lm}). The reason for introducing the model
space $S_0(E_0)$, with splitting the complete Hilbert space $S = S_0 + S'$
and using the LPA in the complementary space $S'$, is as follows.
The subspace $S_0(E_0)$ contains all the two-particle states
$(\lambda, \lambda')$ with both
one-particle energies $\eps_{\lambda}, \eps_{\lambda'}$ being small,
$\eps_{\lambda}, \eps_{\lambda'} < E_0$.
\footnote{In fact, the difference $\eps_{\lambda}{-}\mu$ is small.
Just such differences enter the denominator of eq.(\ref{Blm}) at
$E{=}2\mu$.}
In the complementary
subspace, $S'(E_0)$, one of these energies or both of them are large,
$max(\eps_{\lambda}, \eps_{\lambda'}) > E_0$.
For the model space, the contribution of each
individual state $(\lambda, \lambda')$
\footnote{The "individual" state means the
fixed value of $n,n'$ and small interval of integration over
$k_{\perp}$.}
to the sum of eq.(\ref{Blm}) or eq.(\ref{B1lm})
is strengthened, in comparison with the analogous one for the complementary
space, due to a small value of the energy denominator. Such contributions
induce the long-range terms of the BG propagator $A$ \cite{BLSZ4} and should
be calculated in a direct way. On the contrary, in the complementary
subspace no individual state $(\lambda, \lambda')$ is important and
only wide intervals of the integration over $k_{\perp}$ contribute
to $A$ significantly. The corresponding term of the BG propagator is
sharply peaked and is mainly determined by the local properties
of the system \cite{BLSZ4}. Therefore it is natural to use for it some
kind of local approximation.
For the problem under consideration, it seems to be
more natural to use the LPA rather than the LDA because the BG propagator
in the vicinity of the point $X$ is determined directly by the potential well
$V(X)$ but not by the density, $\rho(X)$. At the same time,
in the surface region there is no simple local relation between
$\rho(X)$ and $V(X)$.

The splitting of the Hilbert space $S = S_0 + S'$ results in the
representation of the BG propagator as sum:
%22
\beq
A=A_0+A',
\label{A01}
\eeq
where $A_0$ contains the states $(\lambda, \lambda')$ which
belong to the model space, $A'$ including the rest.
In accordance with the above consideration, we calculate the
model space propagator $A_0$ explicitly, but use the LPA for
the rest term $A'$. Obviously, the accuracy of the LPA
becomes higher with the model space $S_0$ becoming wider.
We consider the LPA to be good at some value of $E_0$ if the
calculation results for the $G$-matrix do not practically
change with additional increase of $E_0$.

The LPA procedure, in principle, is the same for both
the channels under consideration and is very close to that for the pairing
problem \cite{BLSZ1}, the latter corresponding to the choice $E_0=0$.
Namely, for fixed values of the CM coordinates $X_{12},X_{34}$,
the convolution integral (\ref{Blm}) for $S=0$ (or (\ref{B1lm}), for $S=1$ )
is replaced by the corresponding integral for nuclear matter put in
the constant potential well $V_0=V(X)$, where $X=(X_{12}+X_{34})/2$,
which depends on the difference of the CM coordinates
$t=X_{12}-X_{34}$:
%23
\beq
\Blp_{lm}(X_{12},X_{34};E,{\bf P}_{\perp}) =
\Binf_{lm}(V[X],t;E,{\bf P}_{\perp}).
\label{BLPA}
\eeq

In practice, for a fixed value of the chemical potential $\mu$ and
the cut-off energy $E_0$ and a given set of the potential
depths $V_n$,  we calculate a basic set
$\Binf_{lm}([V_n],t;E=2\mu)$ of nuclear matter propagators.
In fact, we used a sequence of
$V_n{=}\delta V {\cdot} (n-1))$ with the step $\delta V {=}2\;$MeV.
At a fixed coordinate set $X_k$, the elements of the LPA
propagator matrix $\Blp_{lm}(X_i,X_k)$ are found as follows.
First, we find the potential depth $V(X_0{=}(X_i+X_k)/2)$.
Then, for a fixed value of $t=|X_i-X_k|$, the LPA propagator
is found by a linear extrapolation of two neighboring values of
$\Binf_{lm}([V_n],t;E)$,
$\Binf_{lm}([V_{n+1}],t;E)$, under the condition that the inequality
$V_n <V(X_0)<V_{n+1}$ is satisfied.
The convolution integral $\Bfr_{lm}$ for the free propagator
$\Afr$, by definition, coincides with $\Binf_{lm}([V_1=0],t;E)$.
Details can be found in Refs. \cite{BLSZ1}, \cite{BLSZ4}.

\vskip 0.3 cm
              {\bf 4. Validity of the LPA for the singlet channel}
\vskip 0.2 cm

Up to now, we dealt with the general form of the BG equation for
the slab system
containing the total perpendicular momentum $P_{\perp}$ as a
parameter. As it was discussed above, the "dangerous" terms of
the propagator (or of the convolution integrals (\ref{Blm}))
which belong to the model space and should be considered explicitly,
appear due to the small value of the corresponding
denominators in the sum of (\ref{Blm}).
It is obvious that they become more dangerous if
the value of $P_{\perp}$ becomes smaller. Hence
the case of $P_{\perp}=0$ is the most crucial
for validity of the LPA. Therefore we consider just this particular
"bad" case for the LPA. Then, we only focus on
$\mu=-8\;$MeV which is a chemical potential typical of
stable nuclei. Thus, we put in all the
above equations $P_{\perp}=0, E=-16\;$MeV (omitted for brevity
from now on).

As discussed above, the BG equation for the correlation
part of the $G$-matrix (\ref{eqdG}) has a fixed parity $\pi{=}(+,-)$.
Therefore we deal with
eq.(\ref{eqdGpm}) with fixed $\pi$
which is defined only for positive $x$.
After finding the convolution integrals (\ref{Blm})
and those for the free propagator $\Afr$, the kernel of eq.(\ref{eqdG1})
and the nonhomogeneous term (\ref{eqdG2}) are derived by direct
integration. Then one obtains a set of the integral equations for
{\it six} independent components of $\delta G^{\pi}_{ij}(X,X')$
(similar to eq.(\ref{GMeq})) which can be solved numerically
\cite{BLSZ1}, \cite{BLSZ4}. Finally, we find the total correlation
 $\delta G$-matrix (\ref{dGpm}) with components $\delta G_{ij}(X,X')$
or, from eq.(\ref{dG}), the complete $G$-matrix
with components $G_{ij}(X,X')$. They differ by a trivial $\delta$-function
term:
%24
\beq
\delta G_{ij}(X,X')=  G_{ij}(X,X') - \lambda_{ij} \delta (X-X').
\label{dGij}
\eeq
The extraction of the latter makes the quantity under consideration
more convenient for analysis and graphical representation.
Therefore, as a rule, we deal with the only correlation part of the
$G$-matrix, and not with the complete one.
One more remark should be also made before discussing the results.
Following \cite{BLSZ1}, we change the original normalization
\cite{Par1,Par2} of the expansion (\ref{Par1}) in such a way
that the identity $g_i(0)=1$ holds true. In this case, the absolute
values of the $\lambda_{ij}$-coefficients give direct information
on the strength of the corresponding terms of the force. Their values
(in MeV$\cdot$fm$^3$) are as follows:
$\lambda_{11}{=} -3.659 \cdot 10^3 ,\; \lambda_{12} {=}  2.169 \cdot 10^3 ,\;
 \lambda_{22} {=} -1.485 \cdot 10^3 $
 and $\lambda_{13}{=} -2.36 \cdot 10^1 ,\;  \lambda_{23}{=} 5.76 \cdot 10^1
,\; \lambda_{33}{=} 1.72 \cdot 10^1 $.
 The strengths of all the components
containing only the indices $i=1,2$ are
much stronger (by two orders) than those with the index $i=3$. Therefore the
latter are important only for large momenta which come virtually
to the BG equation or the Lippman-Schwinger one. If one analyzes
the matrix elements of the $G$-matrix over the nuclear wave
functions, the typical momenta $k \simeq k_{\Fs}$ appear for
which the contribution of the small components is negligible.
Therefore we, as a rule, concentrate on the "big" components
in a qualitative analysis. Of course,
in the calculations all the terms $\lambda_{ik}$
are taken into account.

We made a series of calculations of the $G$-matrix for several
values of the cut-off energy $E_0{=}0,10,20\;$MeV to analyze
a dependence of the results on this parameter. To present the results
we draw the profile functions $\delta G_{ij}(X,X'{=}X_0)$
of the correlation term of the $G$-matrix at several values
of $X_0$ and the zero moment of the $G$-matrix:
%25
\beq
\bar{G}_{ij}(X) = \int\limits_{-\infty}^{\infty} dt
\,G_{ij} (X,X{+}t).
\label{mom0}
\eeq

A typical example of the profile function of $\delta G_{11}(X,X'{=}0)$
is shown in Fig.~1 for the case of the model space with the
cut-off energy $E_0{=}0$.
It has a sharp peak at the point $X{=}X'$. In such a scale, similar
curves for $E_0{=}10$ and $20\;$MeV are distinguishable from that for
$E_0{=}0$ only after magnification.
Such magnified profile functions for large components with $ij{=}11,12,22$ are
shown in Fig.~2, for $X'{=}0$, and in Fig.~3, for $X'{=}8\;$fm.
It is easily
seen that
already the difference between the curves for $E_0{=}0$ and $E_0{=}10\;$MeV
is rather small. As to that for $E_0{=}10\;$MeV and $E_0{=}20\;$MeV, it looks
negligible.

To analyze the dependence of the $G$-matrix
on $E_0$ in a more quantitative way, it is worth to compare each other
the zero moments (\ref{mom0}) at different values of $E_0$.
They are
shown in Fig.~4 for the same large components and, as an example,
for one small component, $ij{=}13$. One sees that a difference, at a level
of a few percent, exists between the curves for   $E_0{=}0$
and $E_0{=}10\;$MeV and again the additional increase of $E_0$
from 10 MeV  to 20 MeV
does not practically influences the results. This is true not only for
big components, but also for small ones.

Finally, we calculated the "Fermi averaged" $G$-matrix in the $^1S$-channel:
%26
\beq
<{G}_{\Fs}>_{S=0}(X)=
\sum_{ij} \bar {G}_{ij}(X)\,
g_i(k^2_{\Fs}(X))\, g_j(k^2_{\Fs}(X)),
\label{Favr}
\eeq
where we have introduced the local Fermi momentum as
$k_{\Fs}(X){=} \sqrt{2m(\mu{-}V(X))}$
at $\mu{-}V(X)>0$ and which otherwise takes zero value.
Such an average appears if one calculates the Landau-Migdal
amplitude in terms of the $G$-matrix \cite{BLSZ4}.
To this respect, one remark should be made. Though
the profile functions $G_{ij}(X,X')$
are strongly peaked at the point $X=X'$, the long range
"tails", which are hardly seen "by eyes" in Fig.~1,
also contribute to the zero moment (\ref{mom0}).\footnote
{This contribution depends on $ij$ and is, as a rule, not greater than
$10 \div 20$ \%.}
These
terms of the $G$-matrix appear due to states entering the model space
and their contribution to the integral (\ref{mom0}) was analyzed in
\cite{BLSZ4} for the case of $E_0=0$. When one deals with the
problem of evaluation of the Landau-Migdal amplitude which is
supposed to be a short-range coordinate function it is natural
to cut these tales. In Ref.\cite{BLSZ4} a recipe was suggested
to use the Fermi averaged $G$-matrix by eq.(\ref{Favr}) with
the zero moments "with cut-off" which are defined by the integral
of the (\ref{mom0}) type, but with limits of $|t|<t_c$, $t_c{=}3\;$fm.
Of course, for the problem of the validity of the LPA it is not important
what kind of the zero moment is used in (\ref{Favr}).
However, we use here the same recipe for the Fermi averaged $G$-matrix
as in \cite{BLSZ4} because it is more physical.
This quantity is shown in Fig.~5
for the same three values of $E_0$ together with the analogous
average value of the free off-shell $T$-matrix:
%27
\beq
<{T}_{\Fs}>_{S=0}(X)=
\sum_{ij} \bar {T}_{ij}(E{=}2\mu)\,
g_i(k^2_{\Fs}(X))\, g_j(k^2_{\Fs}(X)),
\label{Favrt}
\eeq
where the zero moments $\bar {T}_{ij}$ of the $T$-matrix are defined
similar to (\ref{mom0}). In this case, the introduction of the cut-off
with $t_c{=}3\;$fm does not practically change the integral.
Of course, it is $X$-independent.

Again, the difference between the Fermi-averaged $G$-matrix for
$E_0{=}10\;$MeV and that for $E_0{=}20\;$MeV is negligible.
Their deviation
from the one corresponding to  $E_0{=}0$ is also very small
everywhere except in the surface region. It should be noted that the
difference between the average $G$-matrix and $T$-matrix is
rather small. A similar property was found previously \cite{BLSZ2},
\cite{BLSZ3} for the effective pairing interaction in the $^1S$-channel.

Analysis of Figs.~2 - 5 leads us to the conclusion that for
the singlet channel $S=0$ the LPA works perfectly well for
$E_0{=} 10 \div 20\;$MeV. Moreover, within the accuracy of
a few percent, it is also valid for $E_0{=}0$. The latter
agrees with the analysis of \cite{BLSZ1} where the LPA was
introduced for the pairing problem in the $^1S$-channel.

\vskip 0.3 cm
              {\bf 5. Validity of the LPA for the triplet channel}
\vskip 0.2 cm

In general, the calculation scheme for the triplet $^3S+{^3D}$-channel
is very similar to that for the singlet one, though the calculations become
more cumbersome in this case. Indeed, first, we get {\it ten} independent
components $G_{ij}(X,X')$ and {\it ten}
integral equations (\ref{GMeq}) for them
instead of the {\it six} of the singlet case. Second, the calculation of the
convolution integral (\ref{B1lm}) in the triplet channel is also
more difficult than that of (\ref{Blm}). Therefore the problem of
simplifying these calculations is even more important
than in the singlet channel.

Contrarily to the singlet case, now it is
difficult to separate the multipole terms into the "large" and "small"
ones. Again we changed the original normalization
\cite{Par1} of the expansion (\ref{Par1}),(\ref{Par2})  to guarantee
the identity $g^{L=0}_i(0)=1$
(it should be noted that  $g^{L=2}_i(0)=0$).
Then the strengths of the corresponding terms of the force
(in MeV$\cdot$fm$^3$) are as follows:
$\lambda_{11}{=}-1.618 \times 10^3 ,\;
\lambda_{12} {=}-1.296 \times  10^3  ,\;
  \lambda_{13}{=} 8.921 \times 10^2,\;
\lambda_{14}{=}4.271  \times 10^1 ,\;
\lambda_{22} {=} 7.848 \times  10^2,\;
\lambda_{23}{=} 1.394 \times 10^3,\;
\lambda_{24}{=}-7.860 \times 10^2 ,\;
\lambda_{33} {=} -7.450 \times 10^2,\;
\lambda_{34}{=} -5.723 \times 10^2 ,\;
\lambda_{44}{=} 1.865\times 10^3$.
These values show that, though the strengths of different components
vary significantly, only one of them, $\lambda_{14}$, is less by
two orders of magnitude as compared to the largest ones. Therefore
almost all
the terms are significant. We take several typical components to
illustrate the calculation results.

The profile functions and zero moments are shown in Figs.~6 - 8.
One can see that now the results with increase of the
cut-off energy from $E_0{=}0$ to $E_0{=}10\;$MeV change more
sizably than in the singlet channel, especially in the
surface region.
At the same time,
the subsequent increase of  $E_0$ up to 20 MeV does not practically
influence the $G$-matrix, the maximum variation being of a few percent.
Hence, once more one may conclude that the LPA is sufficiently accurate
if the cut-off energy is $E_0{=}10 \div 20\;$MeV. But, contrarily to
the singlet case, the accuracy of the LPA is rather poor when the model
space is limited to $E_0{=}0$, since at the surface the $G$-matrix must
tend to the off-shell free $T$-matrix. But the latter has a virtual
pole at small energy. It is then clear that an accurate account of
the contribution of the single-particle
states with small positive energies is important for correct description
of this pole behavior. Therefore these states should be included into the
model space $S_0$. This does occur if one chooses the cut-off energy
$E_0 \ge 10\;$MeV, but it does not occur if one takes $E_0{=}0$.

Let us now consider the
Fermi averaged $G$-matrix in the triplet channel, which now is
 a $2 \times 2$ matrix in the orbital angular momentum space:
%28
\beq
<{G}_{\Fs}>^{LL'}_{S{=}1}(X)=
\sum_{ij} \bar {G}_{ij}^{S=1}(X)\,
g^{(L)}_i(k^2_{\Fs}(X))\, g^{(L')}_j(k^2_{\Fs}(X)),
\label{F1avr}
\eeq
where $L,L'{=}0,2$. Just as in the singlet case, the quantity
$\bar {G}_{ij}$ in (\ref{F1avr}) has the meaning of the
zero moment "with cut-off".
The components of this matrix are shown in Fig.~9 for all three values
of the cut-off energy $E_0$.
The component
$<{G}_{\Fs}>^{00}_{S{=}1}$ is significantly larger
than those containing $L{=}2$,
especially at the surface region where the form factors $g_i^{(2)}$
vanish.
Again all the components of the Fermi-averaged $G$-matrix calculated
for $E_0{=}10\;$MeV coincide practically with those for $E_0{=}20\;$MeV,
though deviations from the $E_0{=}0$ case can be noticeable.

\vskip 0.3 cm
              {\bf 6. Conclusion}
\vskip 0.2 cm

The applicability of the LPA has been analyzed for the Brueckner $G$-matrix.
Previously \cite{BLSZ1}
this kind of the local approximation proved to be
quite accurate in the problem of the microscopic
evaluation of the effective pairing interaction in the $^1S$-channel.
The BG equation for a slab of nuclear matter
has been solved for the singlet $^1S$ and  triplet $^3S+{^3D}$ channels
using the separable representation \cite{Par1,Par2} of the Paris potential.
The complete Hilbert space has been split into two domains separated
by the energy $E_0$. The model
subspace $S_0(E_0)$, in which the two-particle BG propagator is calculated
explicitly, contains all the two-particle states with both the
single-particle energies $\eps_{\lambda},\eps_{\lambda'} < E_0$.
In the complementary subspace, $S'(E_0)$,
the LPA for the BG propagator has been used.
A qualitative analysis shows that the accuracy of the LPA becomes
higher with increase of the energy $E_0$. It should be also higher
for larger values of the perpendicular total momentum $P_{\perp}$,
therefore we limit ourselves to the most "dangerous" case of
$P_{\perp}{=}0$.

For either channel under consideration, a
set of calculations of the $G$-matrix has been made for different values
of the cut-off energy $E_0$.
The LPA has been assumed to be
valid starting from the value of $E_0$ for which the $G$-matrix does not
practically change any longer.
An approximate independence of results on the value of $E_0$,
at a level of a few percent, was found for $E_0{=}10 \div 20\;$MeV for both
the channels. It should be mentioned that in the singlet channel
the accuracy of the LPA is sufficiently high even at $E_0{=}0$,
in accordance with \cite{BLSZ1}. On the contrary, in the triplet channel
the LPA is not practically applicable at $E_0{=}0$.

A similar analysis could be made also for the channels with $L>0$.
Estimates show that in this case conditions for validity of the LPA
are even better than those for $L{=}0$.
We believe that the LPA gives a good device for microscopic
describing finite nuclear systems.

In this paper, we limit ourselves to one value of the chemical
potential $\mu{=}-8\;$MeV which is typical for stable nuclei.
In principle, for smaller values of $\mu$ the analysis should be repeated
However, as some estimates show, even in the drip-line vicinity
where $\mu \to 0$ the LPA should be rather good for
$E_0{=}10 \div 20\;$MeV in either channel.
At the same time, at $E_0{=}0$ it should become
inapplicable even in the singlet channel.

\vskip 0.3 cm
              {\bf Acknowledgments}
\vskip 0.2 cm
This research was partially supported by Grant No.~97-0-6.1-7
from the Russian Ministry of Education and Grants
No.~00-15-96590 and No.~00-02-17319 from the Russian
Foundation for Basic Research.
The discussions with S.~A.~Fayans and P.~Schuck are highly acknowledged.

{}

\newpage
\centerline{\bf Figure captions}
\begin{enumerate}

\item %1
The profile function $\delta G_{11}(X,X'=0)$ in the
singlet channel for $E_0=20$ MeV.

\item %2
The profile functions $\delta G_{ij}(X,X'=0)$ in the
singlet channel
for $E_0=0$ (dotted lines), $E_0=10$ MeV (dashed lines),
and $E_0=20$ MeV (solid lines).

\item %3
The profile functions $\delta G_{ij}(X,X'=8)$ in the
singlet channel
for $E_0=0$ (dotted lines), $E_0=10$ MeV (dashed lines),
and $E_0=20$ MeV (solid lines).

\item %4
The zero moments $\bar G_{ij}(X)$ in the singlet channel
for $E_0=0$ (dotted lines), $E_0=10$ MeV (dashed lines),
and $E_0=20$ MeV (solid lines).

\item %5
The Fermi-averaged $G$-matrix in the singlet channel
$<G_{\Fs}>_{S=0}(X)$
for $E_0=0$ (dotted line), $E_0=10$ MeV (dashed line),
and $E_0=20$ MeV (solid line) and the Fermi-averaged
$T$-matrix (thin solid line).

\item %6
The same as in Fig.~2 for the triplet channel.

\item %7
The same as in Fig.~3 for the triplet channel.

\item %8
The same as in Fig.~4 for the triplet channel.

\item %9
The same as in Fig.~5 for the triplet channel.

\end{enumerate}

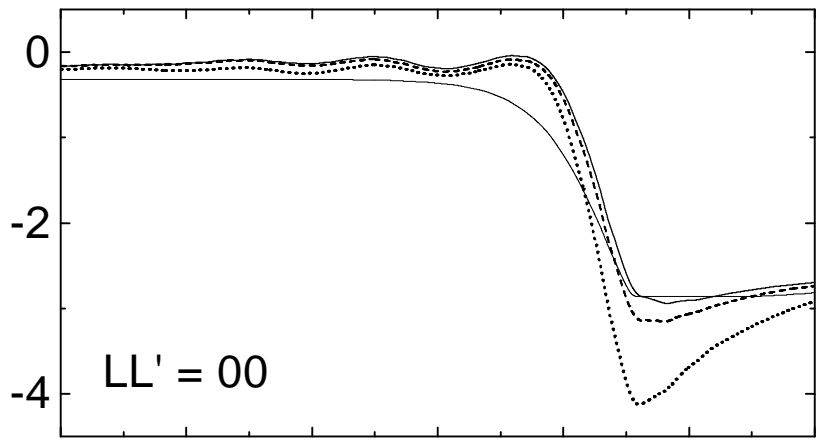
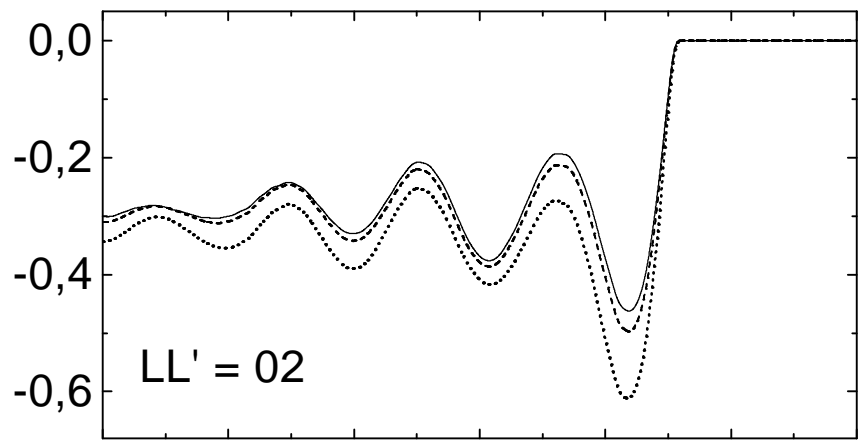
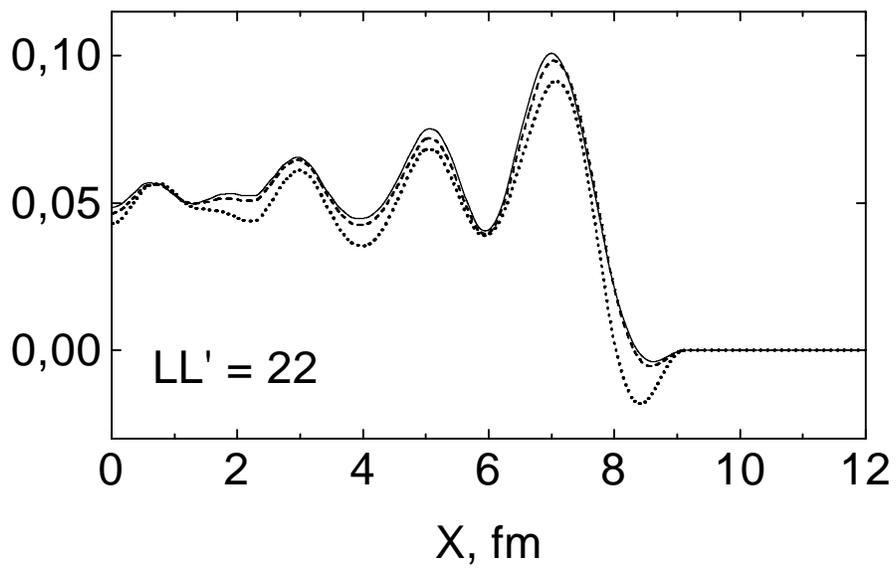
\end{document}